\documentclass[twocolumn,french,english,english,aps,prb]{revtex4}
\usepackage[T1]{fontenc}
\usepackage[latin1]{inputenc}
\usepackage{babel}
\usepackage{graphics}

\makeatletter

\providecommand{\LyX}{L\kern-.1667em\lower.25em\hbox{Y}\kern-.125emX\@}

\usepackage[T1]{fontenc}
\usepackage[latin1]{inputenc}
\usepackage{graphics}

\makeatletter

\usepackage[T1]{fontenc}
\usepackage[latin1]{inputenc}
\usepackage{babel}
\usepackage{graphics}

\makeatletter

\usepackage{amssymb}

\usepackage[dvips]{graphicx}
\usepackage{dcolumn}
\usepackage{amsmath}

\makeatother

\makeatother

\makeatother
\begin{document}

\title{Upper critical field in a trigonal unconventional superconductor
UPt\( _{3} \)}

\author{P. L. Krotkov}

\address{L. D. Landau Institute for Theoretical Physics, Russian Academy of
Sciences, 117334 Moscow, Russia}

\author{V. P. Mineev }

\selectlanguage{french}
\address{D\'{e}partament de Recherche Fondamentale sur la Mati\`{e}re Condens\'{e}e,
Commisariat \`{a} l'\'{E}nergie Atomique, Grenoble 38054, France}
\selectlanguage{english}

\date{\today{}}

\begin{abstract}
A theory of the upper critical field in a trigonal superconductor
with a two-component order-parameter is developed. We demonstrate
the existence of sixfold modulations of the upper critical field in
the basal plane. The form of the angular dependence of \( H_{c2} \)
in the plane of the main symmetry axis is studied in the whole range
of phenomenological rigidity coefficients. A qualitative dependence
of these coefficients on the form of the Fermi surface is established.
The results obtained are discussed in connection with the measurements
of the direction dependence of \( H_{c2} \) in the heavy fermion
superconductor UPt\( _{3} \) that has been recently found to possess
trigonal crystalline structure.

\end{abstract}
\maketitle
\newcommand{\f}[1]{\mbox {\boldmath\(#1\)}}

\newcommand{\TS}{\textstyle}

\newcommand{\DS}{\displaystyle}

\section{Introduction}

The heavy-fermion superconductor UPt\( _{3} \) is traditionally held
to be hexagonal. Based on high-energy x-ray diffraction and transmission
electron microscopy, the crystal structure of UPt\( _{3} \) have
recently \cite{walko} been argued to be determined to be trigonal.

Trigonal deviations from the {}``ideal'' hexagonal symmetry are
of the order of 1\%. That is why one can expect only small changes
in the quantitative interpretation \cite{Graf00} of the numerous
measurements of the thermodynamic and kinetic values based on the
\( E_{2u} \) two-dimensional superconducting state. The same one
can say in respect of the interpretation of the flux-line lattice
experiment \cite{Huxley00} proposed in Ref. \cite{Champel01}. It
seems to be interesting however to concentrate efforts on the study
of the properties that can demonstrate clear indubitable distinction
between the two component superconductivity in hexagonal and trigonal
crystals. The simplest example of such a property is orientational
dependence of the upper critical field.

In the present article we develop Ginzburg-Landau theory of the orientational
dependence of the upper critical field for a two-component superconducting
state in a crystal with a trigonal symmetry. We demonstrate, in particular,
that the upper critical field in a trigonal unconventional superconductor
exhibits six-fold modulations in the basal plane, this conclusion
remaining intact in the presence of an orthorhombic anisotropy in
the direction fixed parallel or perpendicular to the magnetic field.

It is interesting to note that such small modulations (in experiments
\( |\delta H_{c2}^{hex}/H_{c2}|\lesssim 0.02 \)) have been observed
in UPt\( _{3} \) by angular resolved magnetoresistance measurements
\cite{keller94}.

This experimental observation then encountered difficulty in explaining
in the framework of any model of the superconducting phases in a hexagonal
superconductor. Burlachkov \cite{burlachkov85} showed that \( H_{c2}^{\perp } \)
is isotropic for fields in the basal plane near \( T_{c} \) if the
order parameter \( \f \eta  \) transforms according to the two-dimensional
irreducible representation \( E_{1} \) of the hexagonal point group
\( D_{6h} \). He pointed out that \( H_{c2}^{\perp } \) is also
isotropic for \( E_{2} \).

The hexagonal anisotropy in the GL functional then shows up only in
the terms of the sixth order in gradients, giving that the resulting
hexagonal anisotropy of \( H_{c2}^{\perp } \) vanishes as \( (1-T/T_{c})^{3} \)
near \( T_{c} \) just as it does in a hexagonal superconductor described
by a one-component order parameter.

Shortly after the experiments \cite{keller94}, Mineev \cite{mineev94}
demonstrated that in the model of two accidentally nearly degenerate
1D representations the hexagonal anisotropy of \( H_{c2}^{\perp } \)
also vanishes as \( (1-T/T_{c})^{3} \) near \( T_{c} \).

Sauls \cite{sauls96} considered the hexagonal oscillations in the
model of a two-dimensional representation for the superconducting
order parameter coupled to an in-plane antiferromagnetic (AFM) order
parameter \( \mathbf{m}_{s} \). Large Zeeman energy oriented \( \mathbf{m}_{s} \)
near perpendicular to the external magnetic field. Hexagonal anisotropy
of H\( _{c2} \) then arose from the weak in-plane anisotropy energy
of the AFM state through the coupling of the superconducting order
parameter with \( \mathbf{m}_{s} \).

On the other hand, the existence of the six-fold anisotropy \( \propto (1-T/T_{c}) \)
of the upper critical field in the basal plane of a trigonal crystal
is quite natural, as we will see, and does not require any additional
assumptions. It has been noticed for the first time numerically in
the theoretical paper \cite{burlachkov90}, the main goal of which
was to study another type of anisotropy of the upper critical field
--- in the plane passing through the high-symmetry axis of an exotic
trigonal superconducting monocrystal of Cu\( _{1.8} \)Mo\( _{6} \)S\( _{8} \).

We also find the change of the form of \( H_{c2} \) in the plane
of the main symmetry axis with the change of the parameters of the
GL functional placing emphasis on the role of trigonality. We determine
as well the qualitative dependence of the values of these parameters
in the microscopic weak-coupling BCS theory on hexagonal and trigonal
contributions to the shape of the Fermi surface.

The paper is organized as follows. In Sec. \ref{Sec-GLE} we write
out phenomenological GL differential equations on the spatial form
of the order parameter used to find upper critical field in the basal
plane in Sec. \ref{Basal} and in the plane of the main symmetry axis
in Sec. \ref{AC}. The rigidity coefficients in the gradient energy
of the GL functional are calculated microscopically in the weak-coupling
BCS theory as functions of the initial trigonal anisotropies of the
Fermi surface and the order parameter in Sec. \ref{Sec-micro}. Relevance
of the theory to the experimental data known to date is discussed
in the Discussions section \ref{Sec-Disc}.

\section{GL equations \label{Sec-GLE}}

In a trigonal crystal the GL functional up to terms of the second-order
in the gradients and in the order parameter \( \f \eta =(\eta _{1},\eta _{2}) \)
has the form \begin{eqnarray}
\mathcal{F} & = & \alpha |\f \eta |^{2}+\kappa _{1}|D_{i}\eta _{j}|^{2}+\kappa _{2}|D_{i}\eta _{i}|^{2}+\kappa _{3}(D_{i}\eta _{j})(D_{j}\eta _{i})^{*}\nonumber \\
 & + & \kappa _{4}|D_{z}\eta _{i}|^{2}+\kappa _{5}\left( (D_{i}\sigma _{ij}^{k}\eta _{j})(D_{z}\eta _{k})^{*}+\mathrm{c}.\mathrm{c}.\right) ,\label{GLfunc} 
\end{eqnarray}
 where \begin{equation}
\alpha =\alpha _{0}(T-T_{c})/T_{c},
\end{equation}
 the covariant space derivatives \begin{equation}
\label{D}
\f D=-i\f \partial +(2\pi /\Phi _{0})\mathbf{A}
\end{equation}
 substitute the ordinary ones in the presence of the magnetic field
\( \mathbf{H}=\f \partial \times \mathbf{A} \), \( \sigma ^{k} \)
is a vector \( (\sigma ^{z},-\sigma ^{x}) \) of the two Pauli matrices.
\( \Phi _{0}/2\pi =\hbar c/2e \) is the flux quantum.

The condition of the positive definiteness of the functional (\ref{GLfunc})
requires that the following constraints on the coefficients are met
(cf. \cite{BG91})\begin{eqnarray}
\kappa _{1}>|\kappa _{3}|,\quad \kappa _{123} & > & 0,\quad \kappa _{4}>0,\quad \kappa _{123}+\kappa _{2}>0,\nonumber \\
2\kappa ^{2}_{5} & < & \kappa _{13}\kappa _{4},\label{positiveDefiniteness} 
\end{eqnarray}
 where, e.g., \( \kappa _{13} \) denotes \( \kappa _{1}+\kappa _{3} \).

The GL equation \( \delta \mathcal{F}/\delta \f \eta ^{*}=0 \) is
obtained by variation of the energy functional with respect to the
order parameter. Before varying the functional (\ref{GLfunc}) we
note that the vector \( (\sigma ^{z},-\sigma ^{x}) \) of the two
Pauli matrices has the property \( \sigma _{ij}^{k}\eta _{k}=\sigma _{ik}^{j}\eta _{k} \)
which can be checked directly. Using this the term proportional to
\( \kappa _{5} \) in the GL functional can be rewritten as \begin{equation}
\kappa _{5}\left( (D_{i}\sigma _{ij}^{k}\eta _{j})(D_{z}\eta _{k})^{*}+(D_{z}\eta _{j})\sigma _{ij}^{k}(D_{i}\eta _{k})^{*}\right) .
\end{equation}

Performing variation now we get \begin{eqnarray}
-\alpha \f \eta  & = & \left( \kappa _{1}D_{i}^{2}+\kappa _{4}D_{z}^{2}\right) \f \eta +\left( \kappa _{2}\f DD_{i}+\kappa _{3}D_{i}\f D\right) \eta _{i}\nonumber \\
 & + & \kappa _{5}\left( D_{z}D_{i}+D_{i}D_{z}\right) \f \sigma _{ij}\eta _{j}.\label{GLeq} 
\end{eqnarray}

\section{\protect\protect\( H_{c2}\protect \protect \) in the basal plane
\label{Basal}}

For the magnetic field directed along the unit vector \( \hat{\mathbf{n}}=(\cos \varphi ,\sin \varphi ,0) \)
in the basal plane it is convenient to choose the gauge of the vector
potential in the form \( \mathbf{A}=-Hz\widehat{\f \tau } \), where
\( \widehat{\f \tau }=\hat{z}\times \hat{\mathbf{n}} \). Then we
may seek for solutions of equation (\ref{GLeq}) dependent only on
\( z \) and hence put \( \f D=(2\pi /\Phi _{0})\mathbf{A} \) and
\( D_{z}=-i\partial _{z} \) in (\ref{GLeq}). Measuring distances
in units of the length \( \sqrt{\Phi _{0}/2\pi H} \) and denoting
\begin{equation}
\label{lambda}
\lambda =-\alpha \Phi _{0}/2\pi H,
\end{equation}
 we arrive at: \begin{eqnarray}
\lambda \f \eta  & = & \left( -\kappa _{4}\partial _{z}^{2}+\kappa _{1}z^{2}\right) \f \eta +\kappa _{23}z^{2}\widehat{\f \tau }\left( \widehat{\f \tau }\f \eta \right) \nonumber \\
 & + & i\kappa _{5}\hat{\tau }_{i}\f \sigma _{ij}(1+2z\partial _{z})\eta _{j}.\label{GLeq1} 
\end{eqnarray}

Rotating vector \( \f \eta  \) in the basal plane by the angle \( \varphi  \),
i.e. changing variables from \( \eta _{1},\eta _{2} \) to \( \eta _{\parallel }=\hat{\mathbf{n}}\f \eta ,\eta _{\perp }=\widehat{\f \tau }\f \eta  \),
we finally get the system of two second order linear ordinary differential
equations\begin{eqnarray}
\lambda \f \eta  & = & -\kappa _{4}\partial _{z}^{2}\f \eta +\left( \begin{array}{cc}
\kappa _{1} & 0\\
0 & \kappa _{123}
\end{array}\right) z^{2}\f \eta \nonumber \\
 & + & i\kappa _{5}\left( \begin{array}{cc}
-\sin 3\varphi  & -\cos 3\varphi \\
-\cos 3\varphi  & \sin 3\varphi 
\end{array}\right) (1+2z\partial _{z})\f \eta .\label{GLeq2} 
\end{eqnarray}

In a hexagonal superconductor \( \kappa _{5}=0 \) and equations on
\( \eta _{\parallel } \) and \( \eta _{\perp } \) decouple, each
being a Schr\"{o}dinger equation of a harmonic oscillator with its
own rigidity coefficient \begin{eqnarray}
\left( -\kappa _{4}\partial _{z}^{2}+\kappa _{1}z^{2}\right) \eta _{\parallel } & = & \lambda _{\parallel }\eta _{\parallel },\label{Schr} \\
\left( -\kappa _{4}\partial _{z}^{2}+\kappa _{123}z^{2}\right) \eta _{\perp } & = & \lambda _{\perp }\eta _{\perp }.\label{Schr1} 
\end{eqnarray}
 Its eigenvalues with the boundary condition of vanishing at infinity
are\begin{eqnarray}
\lambda _{\parallel } & = & \sqrt{\kappa _{1}\kappa _{4}}(2n+1),\\
\lambda _{\perp } & = & \sqrt{\kappa _{123}\kappa _{4}}(2n+1).
\end{eqnarray}
 When lowering \( H \) at a fixed temperature \( T<T_{c} \) (which
corresponds to raising \( \lambda  \)), the solution realizes for
the first time at the lowest lying level \( \lambda _{n} \) at \( n=0 \).
I.e., a nucleus of the superconducting phase appears for the first
time at\begin{equation}
\label{Hc2parall}
H_{c2\parallel }=\frac{\Phi _{0}}{2\pi }\frac{\alpha _{0}}{\sqrt{\kappa _{1}\kappa _{4}}}\frac{T_{c}-T}{T_{c}}
\end{equation}
 if \( \kappa _{23}>0 \) (and then it is a phase with the order parameter
\( \f \eta  \) parallel to the direction of the applied field), or
at \begin{equation}
\label{Hc2perp}
H_{c2\perp }=\frac{\Phi _{0}}{2\pi }\frac{\alpha _{0}}{\sqrt{\kappa _{123}\kappa _{4}}}\frac{T_{c}-T}{T_{c}}
\end{equation}
 if \( \kappa _{23}<0 \), and then it is a nucleus of the phase with
\( \f \eta \perp \mathbf{H} \). In any case, neither (\ref{Hc2parall})
nor (\ref{Hc2perp}) depends on the direction of the field, so the
upper critical field is rotationally invariant in the basal plane
\cite{burlachkov85}.

In a trigonal crystal the coefficient \( \kappa _{5} \) is non-zero,
and the two equations in (\ref{GLeq1}) cannot be decoupled by a rotation
of the coordinate axes. The influence of the term proportional to
\( \kappa _{5} \) can be taken into account as perturbation.

Unfortunately, this does nor lead to a concise analytical expression.
The reason is that in first order of perturbation theory the correction
to the ground state eigenvalue of the system (\ref{GLeq2}) is given
by the averaging of the perturbation operator \( \propto \kappa _{5} \)
over the unperturbed ground state \( (\psi _{0}(z,\sqrt{\kappa _{1}/\kappa _{4}}),0) \)
or \( (0,\psi _{0}(z,\sqrt{\kappa _{123}/\kappa _{4}})) \) depending
on the sign of \( \kappa _{23} \). Here \begin{equation}
\label{groundState}
\psi _{0}(z,A)=(A/\pi )^{1/4}e^{-Az^{2}/2}
\end{equation}
 is the ground state of the Schrödinger Eq. (\ref{Schr}) or (\ref{Schr1}).
The average of the operator \( 1+2z\partial _{z} \) over the state
(\ref{groundState}) is zero. So one need to account for second order
corrections, which involve an infinite sum of matrix elements of the
operator \( 1+2z\partial _{z} \) between zeroth (\ref{groundState})
and higher lying states. 

One can, however, accomplish perturbational calculations \emph{with
respect to the parameter \( \kappa _{23} \)}, with \( \kappa _{5} \)
generally staying arbitrary. Then one can obtain expressions for the
limit \( \kappa _{5},\kappa _{23}\ll \kappa _{1} \) (see Table I
for values of the coefficients \( \kappa _{23},\kappa _{5} \) in
a BCS theory).

One proceeds by noting that in either of the cases \( \kappa _{23}=0 \)
or \( \kappa _{5}=0 \) by some rotation of the coordinate axes the
two equations (\ref{GLeq1}) decouple. So in both cases the upper
critical field corresponds to the lower of the eigenvalues of the
ground states of the two equations. 

Rotating the coordinate system of Eq. (\ref{GLeq2}) by \( -\pi /4-3\varphi /2 \),
one arrives at\begin{eqnarray}
\lambda \f \eta  & = & \left( -\kappa _{4}\partial _{z}^{2}+\kappa _{1}z^{2}\right) \f \eta +i\kappa _{5}\left( \begin{array}{cc}
1 & 0\\
0 & -1
\end{array}\right) (1+2z\partial _{z})\f \eta \nonumber \\
 & + & \frac{\kappa _{23}}{2}z^{2}\left( \begin{array}{cc}
1+\sin 3\varphi  & -\cos 3\varphi \\
-\cos 3\varphi  & 1-\sin 3\varphi 
\end{array}\right) \f \eta .
\end{eqnarray}

\begin{figure}
{\centering \resizebox*{0.95\columnwidth}{!}{\includegraphics{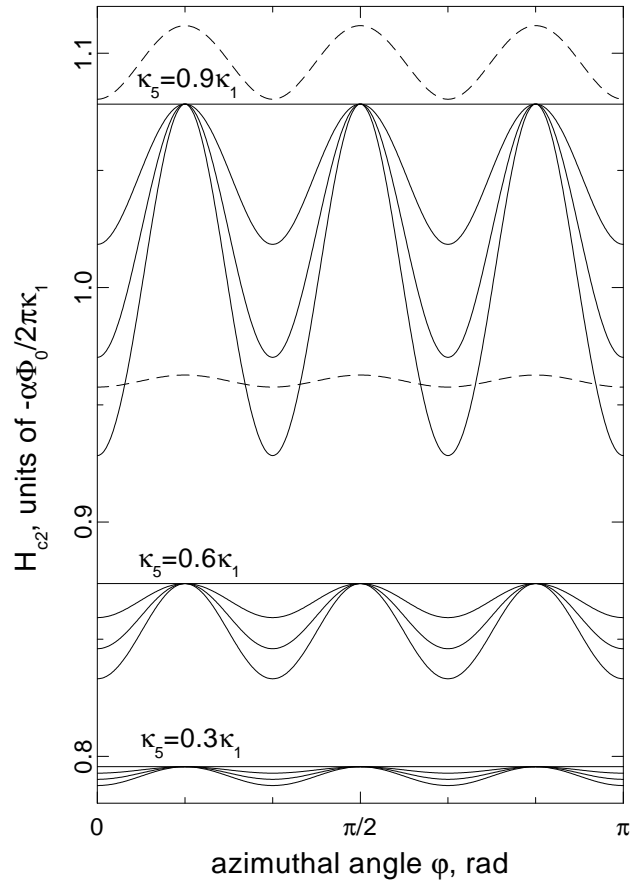}} \par}

\caption{Upper critical field in the basal plane of a two-component trigonal
superconductor for several values of the coefficients \protect\protect\( \kappa _{3}\protect \protect \)
and \protect\protect\( \kappa _{5}\protect \protect \). See explanations
in the text. \label{Fig-basal}}
\end{figure}

In the zeroth order in \( \kappa _{23} \) the two equations decouple
and with the help of the substitution \begin{equation}
\eta _{1}(z)=e^{i\TS \frac{\kappa _{5}}{\kappa _{4}}\frac{z^{2}}{2}}\eta _{1}(z),\quad \eta _{2}(z)=e^{-i\TS \frac{\kappa _{5}}{\kappa _{4}}\frac{z^{2}}{2}}\eta _{2}(z)
\end{equation}
 reduce both to \begin{equation}
\left( -\kappa _{4}\partial _{z}^{2}+\kappa _{1}R^{2}z^{2}\right) \eta =\lambda \eta ,
\end{equation}
 where we introduced the notation \begin{equation}
R=\sqrt{1-\kappa ^{2}_{5}/\kappa _{1}\kappa _{4}}.
\end{equation}
 So that the ground state with the eigenvalue \begin{equation}
\sqrt{\kappa _{1}\kappa _{4}}R\equiv \sqrt{\kappa _{1}\kappa _{4}-\kappa ^{2}_{5}}
\end{equation}
 is twice degenerate with the wave functions \begin{equation}
\left( e^{i\TS \frac{\kappa _{5}}{\kappa _{4}}\frac{z^{2}}{2}}\psi _{0}(z,R\sqrt{\TS \frac{\kappa _{1}}{\kappa _{4}}}),0\right) ,\quad \left( 0,e^{-i\TS \frac{\kappa _{5}}{\kappa _{4}}\frac{z^{2}}{2}}\psi _{0}(z,R\sqrt{\TS \frac{\kappa _{1}}{\kappa _{4}}})\right) .
\end{equation}

Using the first order theory of perturbation of a degenerate level,
we find\begin{equation}
\label{Hc2byK23}
H_{c2}=\frac{H_{c2\parallel }}{R\left[ \DS 1+\frac{\kappa _{23}}{4\kappa _{1}}\frac{1\pm \sqrt{\sin ^{2}3\varphi +R^{3}\cos ^{2}3\varphi }}{R^{2}}\right] },
\end{equation}
 where one should take sign plus for \( \kappa _{23}<0 \) and sign
minus for \( \kappa _{23}>0 \).

We recall that this formula is valid for small \( \kappa _{23}\ll \kappa _{1} \)
and arbitrary \( \kappa _{5} \). For small \( \kappa _{5} \) and
small \( \kappa _{23} \) we get by expansion that the first term
dependent on \( \varphi  \) is \begin{equation}
\label{Ampl}
-\frac{3}{16}H_{c2\parallel }\frac{|\kappa _{23}|}{\kappa _{1}}\frac{\kappa ^{2}_{5}}{\kappa _{1}\kappa _{4}}\cos ^{2}3\varphi 
\end{equation}
 while the constant part of \( H_{c2} \) acquires additional terms
\( O(\kappa _{23}/\kappa _{1}) \) and \( O(\kappa ^{2}_{5}/\kappa _{1}\kappa _{4}) \). 

Eq. (\ref{Ampl}) shows that six-fold modulations of the upper critical
field in a trigonal crystal have very small amplitude for small \( \kappa _{23} \)
and \( \kappa _{5} \), while their maxima occur at \( \varphi =(2n+1)\pi /6 \)
(i.e., at directions of \( \mathbf{H}\parallel \mathbf{a}^{*} \))
irrespective of the signs of \( \kappa _{23} \) and \( \kappa _{5} \).

Numerical calculations confirm these conclusions. They also show that
the dependence \( H_{c2}(\varphi ) \) rests smooth, {}``sinusoidal''
for almost all values of the rigidity coefficients except for both
\( \kappa _{23} \) and \( \kappa _{5} \) large --- each close to
the upper limit of its domain of definition (\ref{positiveDefiniteness}),
when the maxima of \( H_{c2}(\varphi ) \) at \( \varphi =(2n+1)\pi /6 \)
become sharp.

On Fig. \ref{Fig-basal} we presented several curves \( H_{c2}(\varphi ) \)
obtained numerically for \( \kappa _{4}=1.67\kappa _{1} \) and several
\( \kappa _{23} \), \( \kappa _{5} \): 3 groups of solid lines show
curves for \( \kappa _{5}=0.3\kappa _{1} \), 0.6\( \kappa _{1} \),
and 0.9\( \kappa _{1} \) respectively from bottom to top. Inside
each group 4 curves correspond to \( \kappa _{3}=0 \), 0.2\( \kappa _{1} \),
0.5\( \kappa _{1} \), and 0.999\( \kappa _{1} \) (from the upper
line down to the lower). Two dashed lines display \( H_{c2}(\varphi ) \)
for \( \kappa _{3}=-0.15\kappa _{1} \) and \( \kappa _{5}=0.3\kappa _{1} \)
(the lower of the dashed curves) and 0.6\( \kappa _{1} \) (the upper).

An alternative variational method of finding sinusoidal corrections
(\ref{Ampl}) to the upper critical field, which also provides some
insight into the behavior of \( H_{c2} \) in a variant of the GL
theory with an orthorhombic symmetry term, is analyzed in Appendix
A.

\section{\protect\protect\( H_{c2}\protect \protect \) in the plane of the
main symmetry axis \label{AC}}

Upper critical field in the \( \hat{a}\hat{c} \) plane for a trigonal
superconductor was considered in papers \cite{burlachkov90} and \cite{BG91}
in connection with the Chevrel phase compound Cu\( _{1.8} \)Mo\( _{6} \)S\( _{8} \).
We would like to emphasize here the role of the coefficient \( \kappa _{5} \).

In the case of a magnetic field lying in the \( \hat{a}\hat{c} \)
plane the GL differential equation on \( \f \eta  \) analogous to
(\ref{GLeq1}) is more complicated. It can be somewhat simplified
if one assumes\begin{equation}
\kappa _{2}=\kappa _{3}.
\end{equation}
 This equality comes out in the microscopic calculation of the rigidity
coefficients (see Sec. \ref{Sec-micro}), it was also presupposed
in both \cite{burlachkov90} and \cite{BG91}. Conditions (\ref{positiveDefiniteness})
become then \begin{eqnarray}
\kappa _{1}>0,\; \kappa _{4}>0,\; \kappa ^{2}_{5} & < & \frac{\kappa _{13}\kappa _{4}}{2},\; -\frac{\kappa _{1}}{3}<\kappa _{3}<\kappa _{1}.\label{DefInter} 
\end{eqnarray}

However, even with \( \kappa _{2}=\kappa _{3} \) the GL equations
allow analytical solution only for \( \kappa _{2}=\kappa _{3}=0 \),
\( \kappa _{5}=0 \), where the upper critical field \begin{equation}
\label{Hc20}
H^{(0)}_{c2}=-\frac{\Phi _{0}}{2\pi }\frac{\alpha }{\sqrt{\kappa _{1}(\kappa _{1}\cos ^{2}\theta +\kappa _{4}\sin ^{2}\theta )}}
\end{equation}
 turns out to be the same as in the case of a one-component superconductivity
in uniaxial crystals described by the effective mass model (see, e.g.
\cite{Mineev99}).

\begin{figure}
{\centering \resizebox*{0.95\columnwidth}{!}{\includegraphics{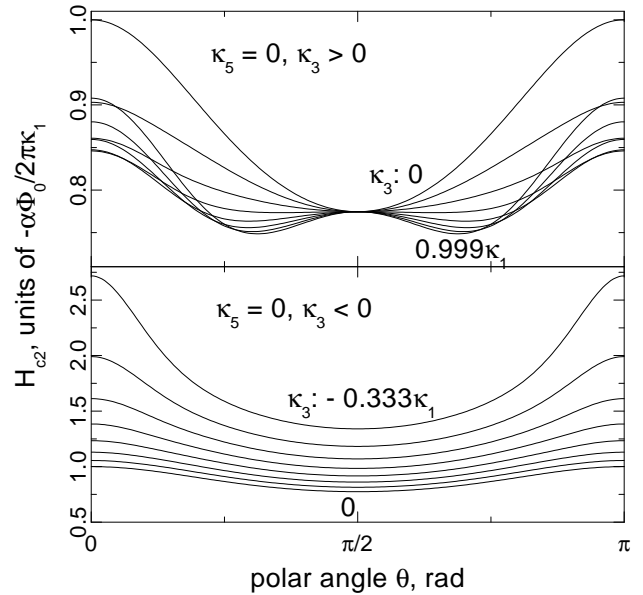}} \par}

\caption{Upper critical field \protect\protect\( H_{c2}\protect \protect \)
in a hexagonal superconductor with two-dimensional order parameter
for a magnetic field lying in the plane of the main symmetry axis
as a function of the angle \protect\protect\( \theta \protect \protect \)
between this axis and the magnetic field. Curves in the two sets correspond
to equally spaced values of \protect\protect\( \kappa _{3}\protect \protect \)
running the interval from 0 to 0.999\protect\protect\( \kappa _{1}\protect \protect \)
(upper plot, from top to bottom) and from 0 to -0.333\protect\protect\( \kappa _{1}\protect \protect \)
(lower plot, from bottom to top). \label{Fig-k50}}
\end{figure}

The ground state of the Schr\"{o}dinger equation whose eigenvalue
is given by (\ref{Hc20}) is twice degenerate. And the influence of
\( \kappa _{3} \) and \( \kappa _{5} \) can be accounted for by
perturbation theory of degenerate levels. Corresponding theory was
constructed in \cite{BG91}. The formulas, though get more accurate
in the second order taking into account crossing of the first-order
levels, are barely citable.

Unexpectedly, a simple first-order expression \begin{eqnarray}
H^{(1)}_{c2} & = & H^{(0)}_{c2}\max \Biggl \{\frac{1}{\DS 1+\frac{\kappa _{3}}{\kappa _{1}}+\frac{\kappa _{5}}{2\widetilde{\kappa }}\sin 2\theta },\nonumber \\
 &  & \frac{1}{\DS 1+\frac{\kappa _{3}}{\widetilde{\kappa }}\cos ^{2}\theta -\frac{\kappa _{5}}{2\widetilde{\kappa }}\sin 2\theta }\Biggr \},\label{Hc21} 
\end{eqnarray}
 where \( H^{(0)}_{c2} \) is given by (\ref{Hc20}) and \begin{equation}
\widetilde{\kappa }=\kappa _{1}\cos ^{2}\theta +\kappa _{4}\sin ^{2}\theta ,
\end{equation}
 is in no much worse accord with the values obtained numerically than
the bulkiest of those in \cite{BG91}.

Note that variational fit with the trial function \( \f \eta =(\eta _{1},\eta _{2})e^{-Az^{2}/2} \)
with three free parameters gives \cite{burlachkov90} an expression
which in the domain of applicability of perturbation theory coincides
with (\ref{Hc21}).
\begin{figure}
{\centering \resizebox*{1\columnwidth}{!}{\includegraphics{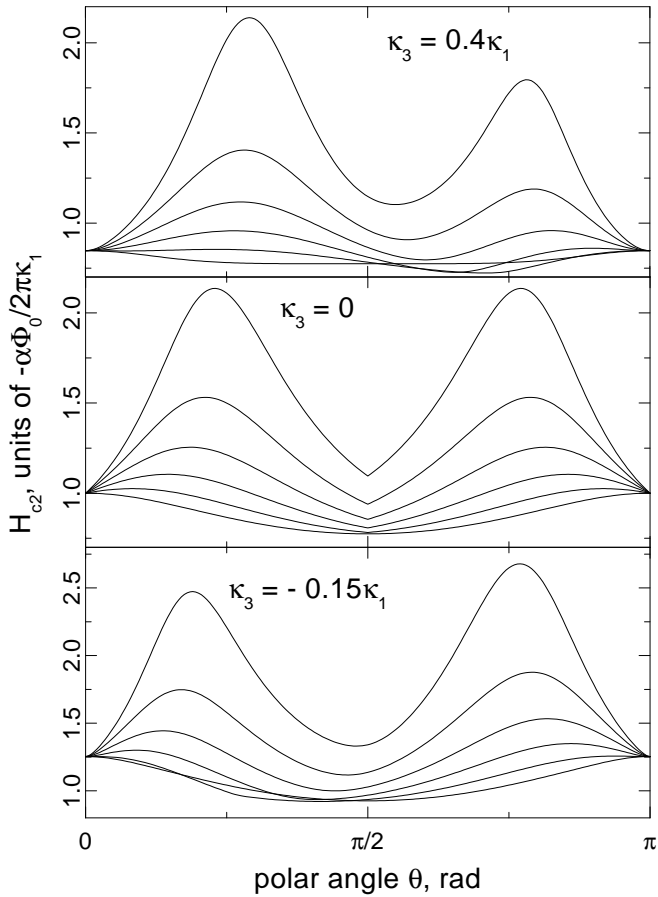}} \par}

\caption{\protect\protect\( H_{c2}\protect \protect \) in a trigonal crystal
(\protect\protect\( \kappa _{5}\ne 0\protect \protect \)) for \protect\protect\( \kappa _{3}=0.4\kappa _{1}\protect \protect \)
(upper plot), \protect\protect\( \kappa _{3}=0\protect \protect \)
(middle plot) and \protect\protect\( \kappa _{3}=-0.15\kappa _{1}\protect \protect \)
(lower plot). In each set of the curves the coefficient \protect\protect\( \kappa _{5}\protect \protect \)
runs over an equally spaced set of values in the interval from 0 (lower
curve in the set) to the upper boundary \protect\protect\( \sqrt{\kappa _{13}\kappa _{4}/2}\protect \protect \)
of the possible values of \protect\protect\( \kappa _{5}\protect \protect \)
(upper curve in the plot).\label{Fig-k5}}
\end{figure}

The dependence of the upper critical field in a hexagonal crystal
(i.e., for \( \kappa _{5}=0 \)) for a magnetic field lying in the
\( \hat{a}\hat{c} \) plane on the angle \( \theta  \) between the
direction of magnetic field and the \( \hat{c} \)-axis is depicted
on Fig. \ref{Fig-k50} for several values of the coefficient \( \kappa _{3} \)
from its definition interval (\ref{DefInter}). \( \kappa _{4}/\kappa _{1} \)
was put equal to 1.67. Only the range of angles from \( 0 \) to \( \pi  \)
is shown because reversing direction of the magnetic field, which
corresponds to the substitution \( \theta \rightarrow \theta +\pi  \),
does not change the value of \( H_{c2} \) resulting from the functional
(\ref{GLfunc}) with (\ref{D}), meaning that the curve on Fig. \ref{Fig-k50}
should be continued periodically.

The curves in each set correspond to the equally spaced values of
the phenomenological coefficient \( \kappa _{3} \) running the range
between \( \kappa _{3}=0 \) and \( \kappa _{3}=0.999\kappa _{1} \)
(upper plot), and between \( \kappa _{3}=0 \) and \( \kappa _{3}=-0.333\kappa _{1} \)
(lower plot) inclusively.

The behavior of \( H_{c2} \) with the increase of \( |\kappa _{3}| \)
is described (for large \( |\kappa _{3}| \) of course only qualitatively)
by (\ref{Hc21}) with \( \kappa _{5} \) put equal to zero. Since
\begin{equation}
\left. \frac{1}{\kappa _{1}}\right/ \frac{\cos ^{2}\theta }{\widetilde{\kappa }}=1+\frac{\kappa _{4}}{\kappa _{1}}\tan ^{2}\theta \geq 1,
\end{equation}
 we conclude that for \( \kappa _{3}<0 \), when the first term in
(\ref{Hc21}) is the larger, the curve of the dependence of \( H_{c2} \)
on \( \theta  \) is scaled with the increase of \( |\kappa _{3}| \)
roughly as \begin{equation}
1/(1+\kappa _{3}/\kappa _{1})
\end{equation}
 without changing its overall shape. Note also that \( H_{c2}(\theta =\frac{\pi }{2}) \)
then equals (\ref{Hc2perp}).

For \( \kappa _{3}>0 \) the upper critical field is described in
the first order of perturbation theory by the second term in (\ref{Hc21}),
and indeed the increase in \( \kappa _{3} \) is accompanied by the
appearance and growing of a {}``bump'' around \( \theta =\pi /2 \).
The value \( H_{c2}(\theta =\frac{\pi }{2}) \) equals (\ref{Hc2parall}).

Non-zero coefficient \( \kappa _{5} \) (trigonality) results in the
discrimination between upper and lower parts of a crystal, and thus
in the curves \( H_{c2}(\theta ) \) non-symmetric with respect to
the point \( \theta =\pi /2 \) if also the coefficient \( \kappa _{3}\ne 0 \)
(see Fig. \ref{Fig-k5}).

At non-zero \( \kappa _{5} \) and \( \kappa _{3} \) the curves are
characterized by the presence of two maxima in the range \( 0<\theta <\pi  \)
with the heights proportional to \( \kappa _{5} \) and the ratio
of the heights proportional to \( |\kappa _{3}| \), though at large
\( |\kappa _{3}| \) the proportionality of the ratio becomes weaker,
so that (\ref{Hc21}) extrapolated to large \( |\kappa _{3}| \) exaggerates
the difference in heights.

A local minimum of \( H_{c2}(\theta ) \) is achieved between the
two maxima at an angle \( \theta <\pi /2 \) for \( \kappa _{3}<0 \)
and \( \theta >\pi /2 \) for \( \kappa _{3}>0 \).

\section{Relative value of the rigidity coefficients \protect\protect\( \kappa _{1},\kappa _{2},\kappa _{3},\kappa _{4}\protect \protect \)
and \protect\protect\( \kappa _{5}\protect \protect \) \label{Sec-micro}}

In the weak-coupling BCS theory these coefficients can be calculated
\cite{sauls94a} as averages over the Fermi surface\begin{equation}
\label{rigiditycoefficients}
\frac{\kappa _{i\mu j\nu }}{\kappa _{0}}=\langle \hat{v}_{\mathrm{F}i}\hat{v}_{\mathrm{F}j}\f \psi _{\mu }(\hat{k})\f \psi _{\nu }(\hat{k})\rangle _{\mathrm{F}},
\end{equation}
 where \( \f \psi _{\mu }(\hat{k}) \) are normalized vector basis
functions (in the \( \mathbf{k} \)-space) of the irreducible representation
\( \Gamma  \) according to which the order parameter \begin{equation}
\mathbf{d}_{\hat{k}}=\sum _{\mu =1}^{\dim \Gamma }\eta _{\mu }\f \psi _{\mu }(\hat{k})
\end{equation}
 transforms.
\begin{figure}
{\centering
\resizebox*{0.9\columnwidth}{0.2\textheight}{\includegraphics{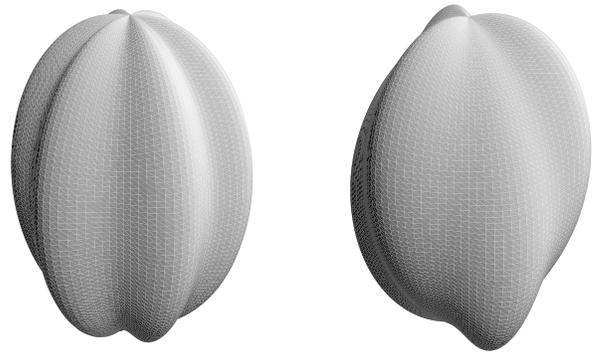}} \par}

\caption{Visualization of hexagonal (left) and trigonal (right) perturbations
to an ellipsoidal Fermi surface used to study the qualitative dependencies
of the values of the rigidity coefficients on the form of the Fermi
surface.\label{Fig-visual}}
\end{figure}

The details of the calculations are placed into Appendix B. Here we
describe the conclusions.

\newcommand{\St}{\raisebox{-2ex}{\rule{0pt}{5ex}}}

\begin{widetext}

\begin{table}[!t]

\caption{Rigidity coefficients for various basis functions and a spherical Fermi-surface
with hexagonal and trigonal perturbations. \label{RigidityCoefficients}}
{\centering \begin{tabular}{rrcrrrcrrrcrrrr}
\hline 
\hline\St Basis \( \psi _{\mu }(\hat{k}) \)&
&
\multicolumn{3}{c}{\( \frac{\kappa _{1}}{\kappa _{0}} \)}&
&
\multicolumn{3}{c}{\( \frac{\kappa _{2}}{\kappa _{0}}=\frac{\kappa _{3}}{\kappa _{0}} \)}&
&
\multicolumn{3}{c}{\( \frac{\kappa _{4}}{\kappa _{0}} \)}&
&
 \( \frac{\kappa _{5}}{\kappa _{0}} \)\\
\hline 
\St\( (\hat{k}_{x},\hat{k}_{y})\hat{\mathbf{z}} \)&
&
 \( \frac{1}{5} \)&
&
 \(  \)&
&
 \( \frac{1}{5} \)&
&
 \(  \)&
&
 \( \frac{1}{5} \)&
&
 \(  \)&
&
 \( \frac{16}{105}\alpha _{\mathrm{tr}} \)\\
 \St\( \hat{k}_{z}(\hat{k}_{x}^{2}-\hat{k}_{y}^{2},-2\hat{k}_{x}\hat{k}_{y})\hat{\mathbf{z}} \)&
&
 \( \frac{1}{3} \)&
\( + \)&
 \( \frac{256}{429}\alpha _{\mathrm{hex}} \)&
&
&
\( - \)&
 \( \frac{256}{429}\alpha _{\mathrm{hex}} \)&
&
 \( \frac{1}{3} \)&
&
 \(  \)&
&
 \( -\frac{68}{429}\alpha _{\mathrm{tr}} \)\\
 \St\( \hat{k}_{z}(\hat{k}_{x}^{3}-3\hat{k}_{x}\hat{k}_{y}^{2})\, \, (\hat{k}_{x},\hat{k}_{y})\hat{\mathbf{z}} \)&
&
 \( \frac{5}{26} \)&
\( + \)&
 \( \frac{56}{4199}\alpha _{\mathrm{hex}} \)&
&
 \( \frac{5}{26} \)&
+&
 \( \frac{56}{4199}\alpha _{\mathrm{hex}} \)&
&
 \( \frac{3}{13} \)&
 \( - \)&
 \( \frac{2352}{4199}\alpha _{\mathrm{hex}} \)&
&
\multicolumn{1}{r}{\( \frac{12}{221}\alpha _{\mathrm{tr}} \)}\\
 \St\( \hat{k}_{z}(\hat{k}_{x}^{3}-3\hat{k}_{x}\hat{k}_{y}^{2})\, \, \hat{k}_{z}(\hat{k}_{x}^{2}-\hat{k}_{y}^{2},-2\hat{k}_{x}\hat{k}_{y})\hat{\mathbf{z}} \)&
&
 \( \frac{9}{34} \)&
\( - \)&
 \( \frac{4032}{7429}\alpha _{\mathrm{hex}} \)&
&
 \( \frac{3}{34} \)&
+&
 \( \frac{4800}{7429}\alpha _{\mathrm{hex}} \)&
&
 \( \frac{5}{17} \)&
 \( - \)&
 \( \frac{4480}{7429}\alpha _{\mathrm{hex}} \)&
&
 \( -\frac{42}{323}\alpha _{\mathrm{tr}} \) \\
\hline 
\hline 
&
&
&
&
&
&
&
&
&
&
&
&
&
&
\\
\end{tabular}\par}\end{table}

\end{widetext}

The symmetry of a crystal can be taken into account in two ways. First,
it is necessarily reflected in the symmetry of the Fermi surface of
a superconductor. And secondly, it should be complied with by the
basis functions \( \f \psi _{\mu }(\hat{k}) \) of the superconducting
state. The shape of the Fermi surface (namely, its cross sections)
is studied in the dHvA experiments. The basis functions are totally
unobservable.

We have modeled the Fermi surface by a sphere (for simplicity of calculation)
with superimposed small hexagonal and trigonal perturbations (see
Fig. \ref{Fig-visual}):\begin{eqnarray}
f(\hat{k}) & = & \alpha _{\mathrm{hex}}\Re (\hat{k}_{x}+i\hat{k}_{y})^{6}+\alpha _{\mathrm{tr}}\hat{k}_{z}\Re (\hat{k}_{x}+i\hat{k}_{y})^{3}\label{formOfPert} \\
 & = & \sin ^{3}\theta \left( \alpha _{\mathrm{hex}}\sin ^{3}\theta \cos 6\varphi +\alpha _{\mathrm{tr}}\cos \theta \sin 3\varphi \right) ,\nonumber 
\end{eqnarray}
 where \( \alpha _{\mathrm{hex}} \) and \( \alpha _{\mathrm{tr}} \)
are the small phenomenological parameters defining the strength of
the perturbation.

\begin{figure}
{\centering \resizebox*{0.6\columnwidth}{!}{\includegraphics{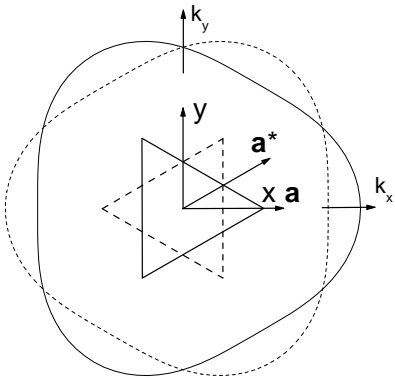}} \par}

\caption{Schematic representation of the orientation of the coordinate system
with respect to the crystalline axes. Also shown is the form (in the
\protect\protect\( \mathbf{k}\protect \protect \)-space) of the trigonal
perturbation to the Fermi-surface.\label{schema-tr}}
\end{figure}

The factor \( \hat{k}_{z}=\cos \theta  \) in the expression for the
trigonal part of the perturbation differentiates between the upper
and lower halves of the Fermi surface. The form of the upper part
of the perturbed surface is schematically shown with the solid curve
on Fig. \ref{schema-tr}, and that of the lower part --- with the
dashed one. The power of \( \sin \theta  \) in front of each term
is of course a question of convenience, it must only be positive to
make the two poles \( \theta =0,\pi  \) non-singular.

We have taken four sample bases \( \psi _{\mu }(\hat{k}) \) of the
irreducible representation \( E_{u} \) of the group \( D_{3d} \).
The results are presented in Table \ref{RigidityCoefficients}. The
complicated fractional values have no physical meaning, --- e.g. for
an oblate unperturbed Fermi surface they would be different. Does
matter the sensitivity of the rigidity coefficients of the superconducting
state with a particular basis to the Fermi surface perturbation of
a given symmetry. 

The first two bases are the most frequent examples for the \( E_{1u} \)
and \( E_{2u} \) irreducible representations respectively of the
group \( D_{6h} \) \cite{sauls94a}. Since both these representations
become \( E_{u} \) in the less symmetric group \( D_{3d} \), their
bases are valid choices.

The second two bases are simply products of the first two by the function
\( \hat{k}_{z}(\hat{k}_{x}^{3}-3\hat{k}_{x}\hat{k}_{y}^{2})=\hat{k}_{z}\Re (\hat{k}_{x}+i\hat{k}_{y})^{3} \).
Since this function is invariant under transformations only from the
\( D_{3d} \) group and not from \( D_{6h} \), these bases are specific
for the \( E_{u} \) representation of the \( D_{3d} \) group.

Table \ref{RigidityCoefficients} shows that all four bases respond
to trigonal perturbation of the Fermi surface, while only the two
last respond to hexagonal perturbation, --- basis \( (\hat{k}_{x},\hat{k}_{y})\hat{\mathbf{z}} \)
does not respond to the latter at all, and for basis \( \hat{k}_{z}(\hat{k}_{x}^{2}-\hat{k}_{y}^{2},-2\hat{k}_{x}\hat{k}_{y})\hat{\mathbf{z}} \)
the coefficient \( \kappa _{4} \) rests unchanged in the linear order.

\section{Discussion\label{Sec-Disc}}

As we have seen, six-fold modulations \( \propto (1-T/T_{c}) \) of
the upper critical field in the basal plane can be naturally explained
in the framework of a usual GL theory in a trigonal unconventional
superconductor. Experimentally observed modulations (\( |\delta H_{c2}^{hex}/H_{c2}|\lesssim 0.02 \))
are of the same order of magnitude as trigonal deviations from the
hexagonal structure (\( \sim 0.01 \)).

The form of the orientational dependence of \( H_{c2} \) as given
by the theory rests fairly accurately sinusoidal in almost all the
domain of the phenomenological parameters of the GL theory that keep
the GL functional positive definite except for the values of both
the parameters \( \kappa _{3} \) and \( \kappa _{5} \) extremely
close to the upper limit of their domains.

At the same time, the curves of the angular dependence of resistivity
having been observed in the transport experiments \cite{keller94}
and their continuation \cite{Rodiere01}, possessed sharp minima at
\( \varphi =\pi n/3 \) (i.e., at directions of \( \mathbf{H}\parallel \mathbf{a} \)).
Not discussing here the legitimacy of mapping form of \( \rho (\varphi ) \)
onto \( H_{c2}(\varphi ) \) we remark here that the observed sharp
dips can't be accounted for by the present theory of equilibrium upper
critical field.

In contrast, the six-fold modulations in the theory developed by Sauls
\cite{sauls96} in the framework of hexagonal structure of UPt\( _{3} \)
originated from hexagonal in-plane anisotropy of the AFM order parameter
\( \mathbf{m}_{s} \) coupled to the superconducting order parameter,
on the supposition that \( \mathbf{m}_{s} \) stays almost orthogonal
to the external magnetic field.

Although the latter assumption contradicts experiment, the theory
managed to account for the change of sign of the modulations with
the passing through the tetracritical point. Corrections from not
exact orthogonality of \( \mathbf{m}_{s} \) to \( \mathbf{H} \)
also rendered hexagonal modulations non-sinusoidal.

We considered a version of our GL theory with an orthorhombic anisotropy
in Appendix A. Upper critical field retained six-fold modulations
if the orthorhombic anisotropy was fixed in the direction parallel
or perpendicular to the magnetic field, the sign of the modulations
changing at the tetracritical point. However, the modulations then
became \( \propto (1-T/T_{c})^{2} \), and the trigonal invariant
\( \propto \kappa _{5} \) in the GL functional (\ref{GLfunc}) then
prevented the intersection of the two curves \( H_{c2}(T) \). 

Thus experimentally irrelevant hypothesis of \( \mathbf{m}_{s} \)
staying perpendicular to \( \mathbf{H} \) leads in both theories
to similar effects, although six-fold modulations of \( H_{c2} \)
themselves stem from completely different sources.

Experiments \cite{Rodiere01} revealed also correlation between the
six-fold modulations of the upper critical field in the basal plane
of some samples of UPt\( _{3} \) and the appearance of a non-symmetric
peak at \( \theta =\pi /2 \) (i.e., at \( \mathbf{H}\perp \hat{c} \))
of \( H_{c2} \) in the plane passing through the main symmetry axis.

Since UPt\( _{3} \) turned out in reality to have trigonal crystal
structure, one \emph{should} expect both six-fold modulations of the
upper critical field in the basal plane and some non-symmetry \( \theta \leftrightarrow \pi -\theta  \)
in the plane of the symmetry axis. But again, the form of the curves
on Fig. \ref{Fig-k5} bears little similarity with usual hexagonal
curves with superimposed non-symmetric peaks at \( \theta =\pi /2 \)
experimentally observed in the resistivity measurements of \cite{Rodiere01}.

An important and not yet fully understood role, however, is played
by the discovered \cite{walko} merohedry of UPt\( _{3} \) --- division
of a sample on a large number of small (\( \sim  \)200-300\AA) domains
with alternating sign of trigonality.

The existence of such domains probably leaves the conclusion of the
modulation of \( H_{c2} \) in the basal plane intact, for so six-fold
a modulation is shared by all the domains irrespective of the signs
of their trigonality. On the other hand, the magnitude of \( H_{c2} \)
when a field is deflected from the basal plane clearly depends on
the direction in which it is deflected --- in other words, a trigonal
monocrystal does not possess the symmetry \( \hat{c}\leftrightarrow -\hat{c} \).

In the investigated samples total numbers of the domains with the
opposite signs of trigonality roughly coincided, so one should not
probably wonder at the fact that the forms of the experimentally observed
\cite{Rodiere01} and theoretically predicted (Fig. \ref{Fig-k5})
dependencies are in poor agreement. Even the strong non-symmetry \( \theta \leftrightarrow \pi -\theta  \)
of the theoretical curves finds only modest correspondence in the
experimental results probably also due to approximately equal number
of merohedral domains with the opposite signs of trigonality.

In the end, a few words are to be said on the microscopically obtained
qualitative dependence of the trigonal GL coefficient \( \kappa _{5} \)
on the shape of the Fermi surface. We found that \( \kappa _{5} \),
which is responsible for all the effects of the trigonal structure
in the theory, is proportional to a trigonal contribution to the model
Fermi surface. The shape resembling right figure on Fig. \ref{Fig-visual}
was never observed to the author's knowledge in dHvA experiments in
UPt\( _{3} \), probably because, as one can note, the cross-section
of such a figure in linear order of the trigonal perturbation coincides
with that of an unperturbed one.

\section{Acknowledgments}

The authors are grateful to P.~Rodiere and A.~Huxley for useful
conversations and discussion of unpublished data and to J.~A.~Sauls
for the interest in the work and helpful comments. One of the authors
(P.~K.) is also grateful to J.~Flouquet for the granted opportunity
to complete the work at the Center for Nuclear Research in Grenoble.

\appendix

\section{Variational method for \protect\( H_{c2}\protect \) in the basal
plane }

Main features of the upper critical filed (\ref{Hc2byK23}) can be
seen on a variational fit with the trial function\begin{equation}
\f \eta =\left( \begin{array}{c}
\eta _{\parallel }e^{-\sqrt{\kappa _{1}/\kappa _{4}}z^{2}/2}\\
\eta _{\perp }e^{-\sqrt{\kappa _{123}/\kappa _{4}}z^{2}/2}
\end{array}\right) ,
\end{equation}
 where \( \eta _{\perp } \) and \( \eta _{\parallel } \) are to
be found from minimization of the GL functional with respect to them.
The result then turns out to be just a truncated sum of the second
order perturbation theory.

Even simpler though for this particular form of the trial function
is \cite{mineev94} to substitute it directly into Eq. (\ref{GLeq2}),
which then becomes an algebraic equation on \( \eta _{\perp } \),
\( \eta _{\parallel } \). From the condition of vanishing of the
corresponding determinant one then gets for the eigenvalue\begin{equation}
(\lambda -\sqrt{\kappa _{1}\kappa _{4}})(\lambda -\sqrt{\kappa _{123}\kappa _{4}})=\sqrt{\kappa _{1}\kappa _{4}}\sqrt{\kappa _{123}\kappa _{4}}A^{2},
\end{equation}
 where \begin{equation}
\label{A2}
A^{2}=\frac{2\kappa ^{2}_{5}}{\kappa _{4}\sqrt[4]{\kappa _{1}\kappa _{123}}}\frac{(\sqrt{\kappa _{1}}-\sqrt{\kappa _{123}})^{2}}{(\sqrt{\kappa _{1}}+\sqrt{\kappa _{123}})^{3}}\cos ^{2}3\varphi .
\end{equation}

Using the definition (\ref{lambda}) of \( \lambda  \), one arrives
at an equation \begin{equation}
\label{EqOnH}
\left( H-H_{c2\parallel }(T)\right) \left( H-H_{c2\perp }(T)\right) =A^{2}H^{2},
\end{equation}
 which describes six-fold modulation of the upper critical field.
E.g., supposed that \( \kappa _{23}>0 \), i.e., that the superconducting
state with the order parameter parallel to the magnetic field arises
first, we find\begin{equation}
\label{six-fold}
H_{c2}\approx H_{c2\parallel }\left( 1+\frac{A^{2}H_{c2\parallel }}{H_{c2\parallel }-H_{c2\perp }}\right) ,
\end{equation}
 or, substituting (\ref{Hc2parall}) and (\ref{Hc2perp}), \begin{equation}
\label{six-fold2}
H_{c2}\approx H_{c2\parallel }\left( 1+2\frac{\kappa ^{2}_{5}}{\kappa _{4}}\sqrt[4]{\frac{\kappa _{123}}{\kappa _{1}}}\frac{\sqrt{\kappa _{123}}-\sqrt{\kappa _{1}}}{(\sqrt{\kappa _{123}}+\sqrt{\kappa _{1}})^{3}}\cos ^{2}3\varphi \right) .
\end{equation}
 For \( \kappa _{23}<0 \) we get an analogous expression with the
opposite sign in front of the corrections.

It is interesting to remark that a six-fold modulation of \( H_{c2} \)
in the basal plane of a trigonal crystal appears also in the presence
of an orthorhombic anisotropy (see, e.g., \cite{Mineev99}, \cite{sauls94a})
in the direction fixed parallel or perpendicular to the magnetic field.
The corresponding term was introduced to the GL theory to account
for the splitting of the superconducting transition in UPt\( _{3} \).

With such a term one of the fields \( H_{c2\parallel }(T) \) or \( H_{c2\perp }(T) \)
becomes \( \propto (1-T/T_{c*}) \) rather than \( \propto (1-T/T_{c}) \),
where the splitting between \( T_{c} \) and \( T_{c*} \) is proportional
to the magnitude of the orthorhombic anisotropy. Eq. (\ref{EqOnH})
is still valid then and produces six-fold corrections (\ref{six-fold})
to \( H_{c2} \), though proportional to \( (1-T/T_{c})^{2} \) unlike
(\ref{six-fold2}), where they are linear. In such a version of the
theory, however, the term \( \propto \kappa _{5} \) in the GL functional
leads to the repulsion of the two lines of \( H_{c2}(T) \) so that
instead of a tetracritical point one has a splitting \( \propto A \).

\section{Calculation of the rigidity coefficients}

Relative values of the rigidity coefficients is convenient to study
with the model dispersion law of electrons \begin{equation}
\label{band-str}
\epsilon _{\mathbf{k}}=\frac{k^{2}}{2m}+2\epsilon _{\mathrm{F}}f(\hat{k}),
\end{equation}
 where \( f(\hat{k}) \) is a small perturbation with a symmetry determined
by the symmetry of the crystal. For simplicity we will consider all
the expressions up to leading term in \( f \). The form of the perturbation
of interest to us is (\ref{formOfPert}) as stated in the main text.

The Fermi-surface is defined by \( \epsilon _{\mathbf{k}}=\epsilon _{\mathrm{F}} \),
whence we obtain its equation in the parameterization of the unit
vector \( \hat{k} \)\begin{equation}
\mathbf{k}(\hat{k})=k_{\mathrm{F}}(1-f(\hat{k}))\hat{k},
\end{equation}
 where \( k_{\mathrm{F}}=\sqrt{2m\epsilon _{\mathrm{F}}} \). Making
use of the value of \( \hat{k} \) in terms of the spherical angles
\( \theta ,\varphi  \) one obtains for the element of the Fermi surface
\begin{equation}
|\partial _{\theta }\mathbf{k}\times \partial _{\varphi }\mathbf{k}|d\theta d\varphi =k_{\mathrm{F}}^{2}(1-2f(\hat{k}))\sin \theta d\theta d\varphi .
\end{equation}

The unit vector \( \hat{v}_{\mathrm{F}} \) of the Fermi velocity
\( \mathbf{v}_{\mathrm{F}}=\left. \f \partial _{\mathbf{k}}\epsilon _{\mathbf{k}}\right| _{\mathrm{F}} \)
can be calculated from (\ref{band-str}) to be \begin{equation}
\hat{v}_{\mathrm{F}}=\hat{k}+\partial _{\hat{k}}f-\hat{k}\left( \hat{k}\partial _{\hat{k}}f\right) .
\end{equation}
 Thus the averaging (\ref{rigiditycoefficients}) over the Fermi-surface
may be written as \begin{eqnarray}
\frac{\kappa _{\mu i\nu j}}{\kappa _{0}} & = & \frac{\langle (1-2f(\hat{k}))\hat{v}_{\mathrm{F}i}\hat{v}_{\mathrm{F}j}\f \psi _{\mu }(\hat{k})\f \psi _{\nu }(\hat{k})\rangle _{\hat{k}}}{\langle 1-2f(\hat{k})\rangle _{\hat{k}}}\label{kappa-tensor} \\
 & = & \left( 1+2\langle f(\hat{k})\rangle _{\hat{k}}\right) \langle \hat{k}_{i}\hat{k}_{j}\f \psi _{\mu }(\hat{k})\f \psi _{\nu }(\hat{k})\rangle _{\hat{k}}\nonumber \\
 & + & \biggl \langle \bigl [\hat{k}_{i}\bigl (\partial _{\hat{k}_{j}}f-\hat{k}_{j}(\hat{k}\partial _{\hat{k}}f)-\hat{k}_{j}f\bigr )\nonumber \\
 & + & \hat{k}_{j}\bigl (\partial _{\hat{k}_{i}}f-\hat{k}_{i}(\hat{k}\partial _{\hat{k}}f)-\hat{k}_{i}f\bigr )\bigr ]\f \psi _{\mu }(\hat{k})\f \psi _{\nu }(\hat{k})\biggr \rangle _{\hat{k}}.\nonumber 
\end{eqnarray}

Calculating the integrals in (\ref{kappa-tensor}) with the bases
functions described in the main text, one arrives at Table~\ref{RigidityCoefficients}.


\begin{thebibliography}{10}
\bibitem{NormOfPsi}For brevity we do not write the normalization factors which should
be such as to \( \langle \f \psi _{\mu }(\hat{k})\f \psi _{\nu }(\hat{k})\rangle _{\hat{k}}=\delta _{\mu \nu } \).
\end{thebibliography}
\end{document}